\documentstyle[prd,aps,floats,epsfig]{revtex}  
\def\beq{\begin{equation}}  
\def\eeq{\end{equation}}  
\def\bea{\begin{eqnarray}}  
\def\eea{\end{eqnarray}}  
\def\p0{\phi_{0}}

\def\l{\left}  
\def\r{\right}  
\def\ds{\displaystyle}  
\def\hc{\displaystyle{\frac{\dot{a}}{a}}}

\begin{document}

\twocolumn[\hsize\textwidth\columnwidth\hsize\csname@twocolumnfalse\endcsname  
\date{\today}  
\author{Rachel Bean and  
Jo\~ao Magueijo}  
\address{Theoretical Physics, The Blackett Laboratory,  
Imperial College, Prince Consort Road, London, SW7 2BZ, U.K.}   
\title{Dilaton-derived quintessence scenario leading naturally to 
the late-time acceleration of the Universe }  
\maketitle  
\begin{abstract}  
Quintessence scenarios provide a simple explanation for the observed  
acceleration of the Universe.  Yet, explaining why acceleration did 
not start a long time ago remains a challenge. The idea 
that the transition from radiation to matter domination played a 
dynamical role in triggering acceleration has been put forward in various 
guises.
We propose a simple dilaton-derived 
quintessence model in which temporary vacuum domination is naturally 
triggered by the radiation to matter transition. In this model 
Einstein's gravity is preserved but quintessence couples non-minimally 
to the cold dark matter, but not to ``visible'' matter. Such couplings
have been attributed to the dilaton  in the low-energy limit 
of string theory beyond tree level. We also show how a cosmological
constant in the string frame translates into  a quintessence-type 
of potential in the atomic frame.
\end{abstract}  


]  
\renewcommand{\thefootnote}{\arabic{footnote}}
\setcounter{footnote}{0} 

Recent astronomical observations of distant supernovae 
light-curves \cite{super,super3}
suggest that the expansion of the universe has recently begun 
to accelerate. This observation has deep theoretical implications.
Accelerated expansion is the hallmark of repulsive gravity, which
according to Einstein's theory of relativity can only be achieved
with extreme forms of matter, such as a cosmological constant $\Lambda$
(the vacuum energy). The measurement of a non-zero cosmological constant
vindicates Einstein's greatest ``blunder'', but leaves cosmology with
severe fine-tuning problems. Normal forms of matter are
diluted by expansion; $\Lambda$ is not. In order to achieve
$\Lambda$ domination nowadays and not before, one has to tune the
initial ratio between vacuum and other forms of energy to about a
part in $10^{130}$ \cite{BT}.

Overall cosmologists would rather set $\Lambda=0$, and hope that other,
less extreme, forms of repulsive matter were behind the observed
acceleration of the Universe. Quintessence \cite{peeb,fr,quint}, 
a scalar field $\phi$ 
endowed with a rolling potential, has become a popular alternative.
Such potentials have appeared variously in the context of Kaluza-Klein, 
super-gravity, and string theories (see \cite{pedro}  Section IIB for 
an excellent review). Quintessence has the desirable property that 
its energy density ``scales'' (i.e. remains at constant fraction) or ``tracks''
the dominant form of matter in the Universe. Deviations from 
scaling eventually develop, following which quintessence starts
behaving like a cosmological constant, leading to the observed acceleration
of the Universe.

However, explaining why acceleration only starts nowadays, some 30
expansion times since the Universe became classical,
still requires that quintessence be fine-tuned, either in the field's 
initial conditions or in the parameters of its Lagrangian 
(see however \cite{andy,us}). In general any theory attempting
to explain the cosmological acceleration has to explain
what is special about the present epoch for acceleration to start now.
We propose that the best explanation
for the coincidence of observed
acceleration nowadays is to associate it with our proximity to the
cosmological transition from radiation to dust domination. This
view was first proposed by Barrow and Magueijo \cite{vsl} in a 
different context. 

In \cite{kappa} Armendariz-Picon, Mukhanov,
and Steinhardt proposed $\kappa$-essence, a quintessence-type 
implementation  of this idea. In such a model scaling is only possible
in the radiation epoch, with $\Lambda$ type of behaviour triggered
by the onset of the matter epoch. This type of behaviour is achieved
with a Lagrangian containing a series of non-linear kinetic terms. 
As the authors themselves recognize, such a model serves to illustrate
a point, rather than to provide the simplest and best motivated 
realization of such a dynamics. The purpose of this letter is to show
that a similar dynamics may be realized in much simpler models,
coincident with dilaton models appearing in the low energy 
limit of string theory beyond tree-level \cite{Dam90,Dam94}.

In non-minimal theories radiation and matter have differing effects on 
the dynamics of the quintessence field. These can be interpreted 
in two alternative ways. 
In one we may depart from Einstein's gravity, and 
couple the field $\phi$ to the Ricci scalar $R$ (possibly in the
form $g(\phi)R$)  in the
gravitational action. This amounts to identifying quintessence with the 
a Brans-Dicke field \cite{bd}. The field $\phi$ will then be driven by $R$
as well as its potential. Recalling that $R=0$ for radiation contributions,  
but $R\propto \rho$, the energy density, for non relativistic matter, we see that the extra 
term could in principle push the 
field off scaling at an epoch close to nowadays, providing an ``$R$- boost'' 
\cite{perrotta}. Simple as this idea might 
be, it does not survive close scrutiny; the $R$-boost is in fact deep in
the radiation epoch.
One must remember that the gravitational equations for such a theory 
are also heavily modified, and indeed the work undertaken by 
\cite{amendola,perrotta,uzan} shows that more fine-tuning, if anything,
is required in order to achieve acceleration nowadays.

Another possibility is to retain Einstein's gravity, but to directly
couple quintessence to the matter fields, via a coupling of the 
form $f(\phi){\cal L}_m$. This corresponds to identifying quintessence
with the Einstein's frame formulation of the dilaton and generate 
field-dependent masses and polarisations. These couplings, and the 
general Brans-Dicke coupling, are related by a conformal transformation but 
usually a simple $f(\phi)$ function is mapped into a complicated $g(\phi)$ 
function and vice-versa. Such couplings are heavily constrained
when applied to the visible matter in the Universe, whether to
photons \cite{carroll}, or to what is usually called baryons \cite{Dam94}.
However, it could be that the dilaton coupled differently to visible
matter and to the dark matter of the Universe. This hypothesis
was suggested in \cite{Dam90}, and allows for large couplings
to be consistent with observations.

We consider the general class of theories with an action, in the Einstein 
conformal frame, given by:
\beq\label{action}
{\cal S} =\int 
d^4 x {\sqrt{-g}} {\left( {R\over 2}
+{\cal L}_\phi
+{\cal L}_V+f(\phi){\cal L}_I \right)}
\eeq
in which $8\pi G=1$, ${\cal L}_V$ is the Lagrangian of ``visible matter''
(baryons, photons, and also baryonic and neutrino dark matter),
and ${\cal L}_I$ the Lagrangian of a dominant non-baryonic form
of cold dark matter. As usual 
$
{\cal L}_\phi=-\partial_\mu\phi\partial^\mu\phi/2-V(\phi)
$
with  $V(\phi)=V_0e^{-\lambda\phi}$ the standard
quintessence potential. 
This theory clearly has the potential to behave in line
with the dynamics sought - since it drives quintessence via invisible
matter. In the radiation epoch invisible matter becomes subdominant, 
and we may expect the usual scaling solutions to be valid. In the vicinity
of the transition to matter domination, the new driving term becomes
significant and may induce deviations from scaling. 

Actions with different couplings to each individual matter terms
arise in full-loop expansion generalisations of an effective 
action for the massless modes of a dilaton, for example as 
considered by Damour and Polyakov \cite{Dam94}. These give
an action of the form
\bea
{\cal S} &=&\int 
d^4 x \sqrt{-{\hat g}}  \{{\hat B}_g(\Phi)({\hat R}/2-2{\hat \Lambda})
-{\hat B}_\Phi(\Phi)\partial_\mu\Phi\partial^\mu\Phi\nonumber \\
&& +\sum_i{\hat B}_{(i)}(\Phi){\hat{\cal L}}^{(i)}\}
\eea
where (i) represent the different matter terms, and ${\hat \Lambda}$
is a string frame cosmological constant. 
In \cite{Dam94} it was hoped that the couplings are not too
different for different types of matter, so as not to conflict
with the E\"{o}tvos experiment; however they could be very
different for 
the dark matter of the Universe \cite{Dam90,bellido}. 
A further rationale for why this could be the case is that the dark matter 
may indeed be very exotic (e.g. super-symmetric dark matter),
in which case we may expect the couplings to the dilaton to be 
very different than to ordinary matter.  

Hence, we follow \cite{Dam94}
assuming a Universal coupling $B(\Phi)$ for gravity and all forms
of visible matter, but follow \cite{Dam90} taking the coupling to
invisible matter to have a different strength. For example, the 
higher-order loop corrections to the string coupling could be 
non-negligible giving a coupling of the form \cite{Dam94} 
\beq 
B_I(\Phi)=e^{-2\Phi}+c_{0}+c_{1}e^{2\Phi}+c_{2}e^{4\Phi}+... 
\eeq
with $c_i\neq 0$ parameterizing the corrections beyond tree-level. 
Hence the action can be written 
\bea
{\cal S} &=&\int 
d^4 x {\sqrt{-{\hat g}}}  \{\sigma ({\hat R}/2-2{\hat \Lambda}
+{\hat{\cal L}}_V)
-(\omega/\sigma) \partial_\mu\sigma\partial^\mu\sigma\nonumber\\
&&+{B}_I(\sigma){\hat{\cal L}}_I \}
\eea
where $\sigma={\hat B}(\Phi)$,  as defined in \cite{Dam94}.

Conformally  transforming from the string frame to the Einstein frame 
we obtain the proposed action (\ref{action}), where 
the function $f(\phi)$ can be expressed in terms of the coupling 
$B_I(\Phi)$. The relevant transformation is $g_{\mu\nu}=2\sigma
{\hat g}_{\mu\nu}$ and $2\sigma =e^{-\lambda\phi}$ with
$\lambda=(\omega+3/2)^{1/2}$. 
We highlight the remarkable fact that 
a dilaton independent cosmological constant in the string frame
is transformed into a quintessence potential $V(\phi)={\hat \Lambda}
e^{-\lambda\phi}$
in the Einstein frame \cite{note}.
Hence, the presence of a cosmological constant
in the string frame allows one to identify
the dilaton with the quintessence field.
Note that the Einstein frame for our model is 
identical with the Jordan or atomic frame for visible matter,
in which it follows geodesics; this is usually considered the
physical frame \cite{Dam90}.

The coupling $f(\phi)$ (and also all the $B(\Phi)$) are expected to be 
approaching a minimum \cite{Dam94,arkady}
characterised by $\phi = \phi_{0}$, say.
Hence, for our purposes, the function $f(\phi)$  may be approximated as 
a Taylor expansion about the minimum, 
\beq
f(\phi) = 1+\sum\ds{\frac{1}{\beta!}
\l.\frac{\partial^{\beta}f}{\partial\phi^{\beta}}\r|_{\phi=\phi_{0}}} 
(\phi-\phi_{0})^{\beta} 
\eeq
We therefore investigate a coupling of the form 
$f(\phi)=1+\alpha(\phi-\phi_{0})^{\beta}$ where $\alpha$ and
$\beta$ reflect the concavity of the minimum. 

Varying  action (\ref{action}) with respect to  
the metric and $\phi$ we obtain the field equations: 
\bea 
G_{\mu\nu}&=&T^V_{\mu\nu}+T^\phi_{\mu\nu}+f(\phi)T^I_{\mu\nu}\\ 
\Box \phi&=&{\partial V\over \partial \phi}-{\partial f
\over \partial \phi}{\cal L}_I
\eea 
where $G_{\mu\nu}$ is the Einstein's tensor and the various $T_{\mu\nu}$ 
are stress-energy tensors.
Heuristically, we may interpret the new term driving $\phi$
as a contribution to an effective potential $V_{eff}=V-f(\phi){\cal L}_I$. 
Bianchi's identities ($\nabla_\mu G^\mu_\nu=0$) lead to
integrability conditions: 
\bea 
\nabla_\nu T_V^{\mu\nu}&=&0\\ 
\nabla_\nu T_I^{\mu\nu}&=& (g^{\mu\nu}{\cal L}_I
-T^{\mu\nu}_I) 
{f'\over f}\nabla _\mu\phi  
\eea 
to be contrasted with Amendola's coupled quintessence
\cite{amendola1} (for which the interaction term 
is proportional to $T$). 
 
Interestingly, the equations of motion depend 
on the Lagrangian, and so full divergences are
no longer irrelevant leading to a wealth of possibilities.
For a perfect fluid 
we may infer the Lagrangian from its constituent particles (providing
they do not interact). For a pressureless fluid each particle 
has Lagrangian  
\beq
{\cal L}(x)=-\int d\lambda E_0{\delta(x-y(\lambda))\over {\sqrt {-g}}} 
\sqrt {-g_{\mu\nu}{dy^\mu\over d\lambda}{dy^\nu\over d\lambda}} 
\eeq
where $\lambda$ is the affine parameter (or proper time),
$y(\lambda)$ is the particle's trajectory, and $E_0$ its 
rest mass. Hence we have that for a homogeneous pressureless 
fluid ${\cal L}=-\rho$.  A similar argument applied to relativistic particles
leads to ${\cal L}=0$ for radiation fluids. 
 
Specializing to a flat Friedmann model, with scale 
factor $a$, we find Friedmann equations: 
\bea
3{\left(\dot a \over a\right)} ^2 &=&
\rho_{b}+\rho_{r}+f(\phi)\rho_I+\frac{1}{2}\dot{\phi}^{2}+V(\phi) 
\\ 
\dot{\rho}_I+3\hc\rho_I&=&-{f'(\phi)\dot{\phi}\over f(\phi)}
(\rho_I+{\cal L}_I)=0  \\ 
\rho_b+3\hc\rho_{b}&=&0 \\
\rho_{r}+4\hc\rho_{r}&=&0 \\ 
\ddot{\phi}+3\hc\dot{\phi}+V'&=&f'(\phi){\cal L}_I=-f'(\phi)\rho_I 
\eea 
where dots represent derivatives with 
respect to proper time, and the prime (') indicates differentiation  
with respect to $\phi$.  
 
\begin{figure} 
\centerline{\psfig{file=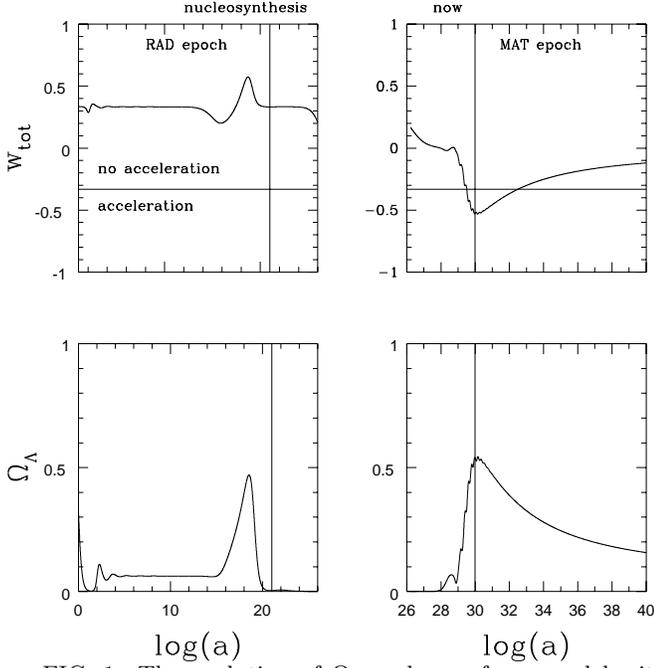,width=9 cm,angle=0}} 
\caption{The evolution of $\Omega_\phi$ and $w_{tot}$ for a model
with $\lambda=8$, $\beta = 8$ , $\alpha =50$, and $\phi_0=32.5$. 
An early period of scaling is broken near the transition from radiation
to matter, first with a period of kination, then inflation. 
At late times the universe returns to a matter dominated 
scaling solution.}
\label{fig1} 
\end{figure}

\begin{figure} 
\centerline{\psfig{file=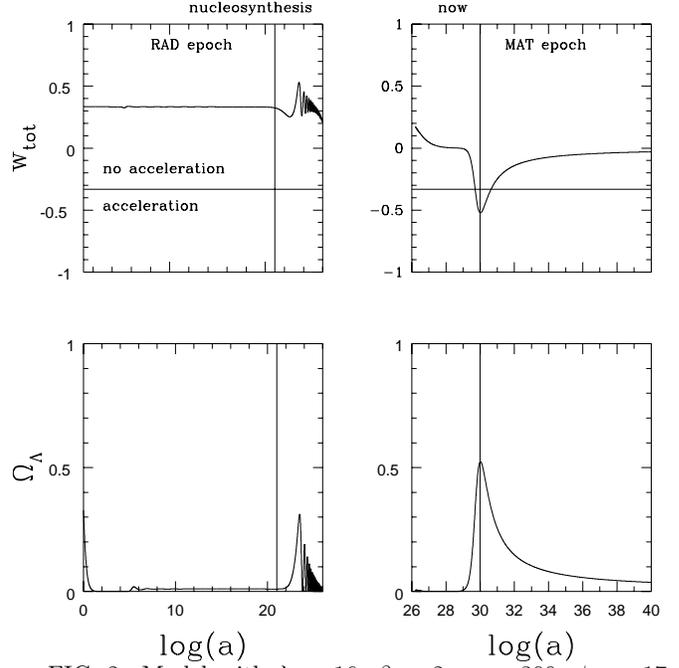,width=9 cm,angle=0}} 
\caption{Model with $\lambda=16$, $\beta= 2$,
$\alpha=300$, $\phi_0 =17$. Notice  the structure of transients
occurring around kination and inflation. Unusually,
in this model kination happens after nucleosynthesis.} 
\label{fig2} 
\end{figure}

In Figs.~\ref{fig1} and \ref{fig2} we plot two typical 
examples of solutions for the cosmological evolution in this theory. 
We plot the fraction of energy in quintessence
$\Omega_\phi=\rho_\phi/\rho_{tot}$, 
and the total equation of state $w_{tot}=p_{tot}/\rho_{tot}$
where $p_{tot}$ is the total pressure (induced by the radiation and
$\phi$). We separate the radiation from the matter
epoch (left and right panels), and indicate where nucleosynthesis
and nowadays lie. 
We see that the driving term $f(\phi){\cal L}_I$, in the form proposed, 
can indeed kick $\phi$ off scaling in the vicinity of $a_{eq}$ 
with a transient regime lasting  4 expansion times after 
{\it and  before} $a_{eq}$.
Typically the field is first pushed into kination to re-emmerge 
into inflationary behaviour, the two events arranging themselves 
symmetrically around $a_{eq}$ along the $\log (a)$ axis.  
 
The acceleration produced in this model is always 
a transient phenomenon. The field $\phi$ produces 
inflation because the driving term $f(\phi)\rho_I$
induces a local minimum in the effective potential 
$V_{eff}=V+f\rho_I$ similar to the one in the
potential proposed in \cite{andy}. However, 
as soon as inflation starts, $\rho_I$ is diluted, 
which in turn withdraws the extra driving force, leading the 
field back into scaling. As in \cite{us}, the observed  
spell of vacuum domination turns out to be a bluff,  
with a new matter epoch following the present $\Lambda$ 
dominance.  This complex feedback process
explains the fast oscillations preceding kination and inflation
for some of the parameters of our model, such as the one
in Fig.~\ref{fig2}. 
  
\begin{figure} 
\centerline{\psfig{file=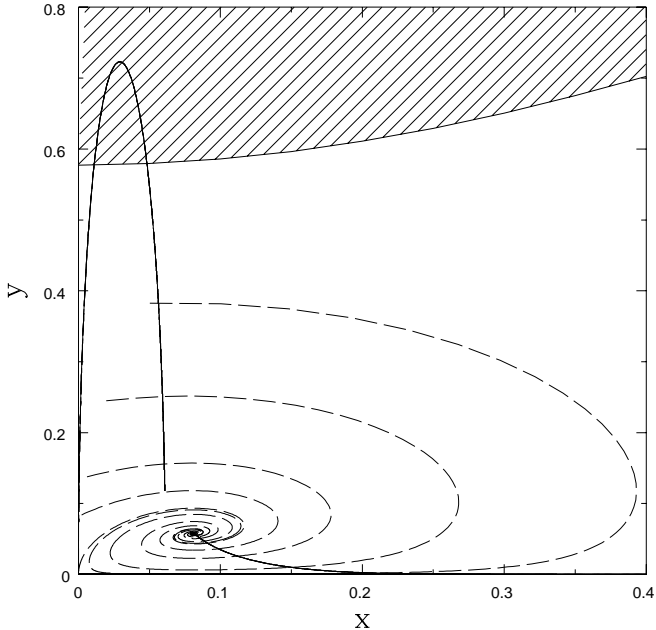,width=9 cm,angle=0}} 
\caption{ Phase space portrait of the model of Fig.2, with
$x=\dot\phi/H 6^{1/2}$ and $y=V^{1/2}/3^{1/2}H$ (with
$H={\dot a}/a$).  Different initial conditions lead to different
orbits. All converge on a fixed point - the radiation epoch 
scaling attractor. Near the radiation to matter transition,
a kination transient pushes all orbits towards the $x$ axis, 
and then up into the shaded inflationary region, before the 
matter dominated scaling fixed point is achieved.}
\label{fig3} 
\end{figure}

Our model illustrates the point that we do not need to have 
an inflationary attractor to explain the current acceleration
of the Universe. Indeed, as shown in Fig.\ref{fig3}
the structure of attractors in our model
is the same as in standard quintessence. It is the motion of the
system while moving between the two (matter and radiation)
attractors  which is new. Perhaps similar transient behaviour 
is present in some extended quintessence models; most of the work done 
so far has focused on attractors \cite{amendola,perrotta,uzan}.

We remark that the symmetry of kination and inflation 
around $a_{eq}$ means that this model bypasses the nucleosynthesis 
constraints usually affecting standard quintessence 
\cite{pedro}. This is because, coincidentally,
nucleosynthesis, equality, and nowadays are roughly equally 
spaced along the  $\log (a)$ axis. Hence, typically kination occurs 
before nucleosynthesis if we want the field to inflate nowadays. 
This means that $\Omega_\phi\approx 0$ during nucleosynthesis, 
invalidating the bound $\lambda> 5$ derived in \cite{pedro}.

However, there are further constraints on this type of model, due 
to the fact that for many purposes it is $f\rho_I$ what should be 
regarded as the matter density (since this is the gravitational mass 
of the invisible matter) and not $\rho_I$ (which is the conserved 
mass). Deciding between the two is mostly a matter of language,  
dependent on whether to count $(f-1){\cal L}_I$ as an interaction 
term or not. In any case, the transition between a radiation epoch 
(with $a\propto t^{1/2}$) and a matter epoch (with $a\propto t^{2/3}$) 
is determined by the redshift for which $\rho_b+\rho_I f =\rho_r $ 
and so is affected by the change in $f$. In general this pushes up 
the redshift of equality, since $f$ is a decreasing function. A 
competing factor results from the reduction of the amount of $\rho_I$ 
nowadays resulting from the current dominance of quintessence.  
This tends to reduce the equality redshift. The first effect is  
normally larger than the second, but can be made arbitrarily small 
by increasing $\lambda$ - so that the change in $\phi$ and $f(\phi)$  
is smaller.  

More important still is the effect such a coupling may have on
the growth of dark matter perturbations. It can be proved
that, in the limit in which fluctuations in $\phi$ are ignored,
the equations for $\delta_I=\delta\rho_I/\rho_I$
are unaffected. Hence, we expect the only effect on the matter
power spectrum to result from the change in the redshift
at equality. No obvious disastrous effect is present;
and so the model we have proposed is not a priori inconsistent
with observations of large scale structure. 
Nonetheless, more subtle effects are present due to 
fluctuations in $\phi$, which  induce new terms in the equations for 
$\delta_I$ (cf. \cite{amendola2}), besides sourcing the perturbed potentials. 
A complete study of structure formation for our model is deferred
to future work, but we suspect results not dissimilar to those
found in standard quintessence scenarios \cite{pedro1}.

In summary, we have found a bridge between dilaton and quintessence
models, by noting that a string frame ${\hat \Lambda}$ transforms into
a rolling potential for the dilaton in the physical frame. The
dilaton may couple with different strengths to visible and dark matter,
a property we used to naturally trigger (transient) acceleration nowadays.
The model is consistent with obvious constraints, but a careful
study of its more subtle effects on structure formation is warranted.

{\bf Acknowledgements} We would like to thank Luca Amendola,
John Barrow, Michael Joyce, and Kelly Stelle for discussions. RB thanks the 
PPARC for support.


\begin{thebibliography}{99}
\bibitem{note}A further transformation $\phi\rightarrow -\phi$
is required to bring the potential to its standard form.
\end{thebibliography}
\end{document}